\documentclass[12pt,preprint]{aastex}
\usepackage{epsf}
\usepackage{epsfig}
\usepackage{psfrag}
\usepackage{changebar}
\usepackage{amsmath}
\usepackage{amssymb}
\def\p{\partial}

\def\l{\left}
\def\r{\right}

\newcommand \dd[2] {\frac{d {#1}}{d {#2}}}
\newcommand \pp[2] {\frac{\p {#1}}{\p {#2}}}
\newcommand \ppd[2] {\frac{\p^2 {#1}}{\p {#2}^2}}

\numberwithin{equation}{section}
\received{2003 August 23}
\begin{document}
%%%%%%%%%%%%%%%%%%Title, author and affiliation%%%%%%%%%%%%%%%
\title{{\bf Gravitational Binding, Virialization and the Peculiar
Velocity Distribution of the Galaxies}}
\date{}
\author{Bernard Leong\altaffilmark{1}}
\affil{Astrophysics Group, Cavendish Laboratory, Madingley Road, 
Cambridge CB3 0HE, England \\
Department of Bioinformatics, Sanger Institute, Hinxton Hall, 
Cambridge CB10 1SA, England }
\and
\author{William C. Saslaw}
\affil{Institute of Astronomy, Cambridge, England \\
Astronomy Department, University of Virginia, Charlottesville, VA\\
National Radio Astronomy Observatory\altaffilmark{2}, Charlottesville, VA}

\altaffiltext{1}{Email to cwbl2@mrao.cam.ac.uk or bcl@sanger.ac.uk}
\altaffiltext{2}{Operated by Associated Universities, Inc., under cooperative
agreement with the National Science Foundation} 
%%%%%%%%%%%%%%%%%%%%%%%%%%%%%%%%%%%%%%%%%%%%%%%%%%%%%%%%%%%%%
%%%%%%%%%%%%%%Abstract%%%%%%%%%%%%%%%%%%%%%%%%%%%%%%%%%%%%%%%
\begin{abstract}
We examine the peculiar velocity distribution function of galaxies
in cosmological many-body gravitational clustering. Our statistical
mechanical approach derives a previous basic assumption and generalizes
earlier results to galaxies with haloes. Comparison with the observed
peculiar velocity distributions indicates that individual massive 
galaxies are usually surrounded by their own haloes, rather than being embedded in
common haloes. We then derive the density of energy states, giving the
probability that a randomly chosen configuration of $N$
galaxies in space is bound and virialized. Gravitational clustering is
very efficient. The results agree well with
the observed probabilities for finding nearby groups containing $N$ galaxies. A
consequence is that our local relatively low mass group is quite
typical, and the observed small departures from the local Hubble flow
beyond our group are highly probable.  
\end{abstract}
%%%%%%%%%%%%%%%%%%%%%%%%%%%%%%%%%%%%%%%%%%%%%%%%%%%%%%%%%%%%%
%%%%keywords%%%%%%%%%%%%%%%%%%%%%%%%%%%%%%%%%%%%%%%%%%%%%%%%%%%%%
%%%%%%%%%%%%%%%%%%%%%%%%%%%%%%%%%%%%%%%%%%%%%%%%%%%%%%%%%%%%%%%%%%
%\doublespace
%%%%%%%%%%%%%%%%%%%%%%%%%%%%%%%%%%%%%%%%%%%%%%%%%%%%%%%%%%%%%%%%%%
\section{INTRODUCTION} \label{intro}
%%%%%%%%%%%%%%%%%%%%%%%%%%%%%%%%%%%%%%%%%%%%%%%%%%%%%%%%%%%%%%%%%%

A fundamental question about any spatial configuration of galaxies is
whether it forms a gravitationally bound group or cluster. If so, we
can ask whether it is sufficiently relaxed to satisfy the virial
theorem, $2 \langle K \rangle +  \langle W \rangle = 0$, relating the
time averages of its kinetic and potential energies. A closely related
problem is to estimate the effects of configurations on the
peculiar velocities of galaxies around them. 

Approximate answers to these questions are known for computer simulations
\citep{saslaw3, saslaw2} and some observed clusters \citep{barychev}. 
In this paper, however, we are concerned with statistical properties of many configurations rather
than individual cases. This will help determine whether the local
group, say, is a typical or unusual configuration. Therefore we will
explore probability distributions for peculiar velocities, total
energies and virialization in regions of gravitational
clustering. Since the observed clustering is highly non-linear on
small scales, our theoretical description will also be fundamentally
non-linear. 

The peculiar velocities of galaxies, i.e. their departures from the
global cosmic expansion, provide a basis for our discussion of binding
and virialization, and contain significant information about the
history of galaxy clustering and the geography of dark matter. Much of
this information can be represented by the velocity distribution
function $f(v) d^3 v$, which is the probability that a galaxy has a
peculiar velocity between $v$ and $v+dv$. In the case of a perfect
gas, this is the familiar Maxwell-Boltzmann distribution. In the case
of galaxy clustering, which is highly non-linear on scales less than
several megaparsecs, $f(v)~d^3 v$ departs greatly from the
Maxwell-Boltzmann form. This is because it includes galaxies with all
degrees of clustering, from isolated field galaxies to the densest
rich clusters, in a very large representative volume of space.

The present strongly non-linear velocity distribution presumably developed from
more quiescent initial conditions. Initial states which start with a linear
Gaussian distribution of density and velocity perturbations can be followed
reasonably well into the weakly non-linear regime using perturbation theory 
\citep{nusser, nusser2, bernardeau}.  On scales where the distributions have evolved 
into strongly non-linear systems, dynamic dissipation has destroyed most of the detailed 
memory of the initial state. For the cosmological many-body case, this relaxation on small spatial
scales, where the gravitational field is very grainy \citep{saslaw10}, takes only one or two
expansion timescales. These non-linearities then spread to larger scales as the
clustering evolves.

The observed velocity distribution function was discovered \citep{raychaudhury}
using a representative sample of galaxies from the Matthewson survey
which has relatively little a priori bias regarding their degree of
clustering. Previously it had been predicted \citep{saslaw10} using a
quasi-equilibrium gravitational many-body theory of galaxy
clustering. This theory in which galaxies may be surrounded by
individual haloes, is in excellent agreement with relevant N-body
simulations and agrees very well with the observations \citep{crane,
fang, sheth, saslaw12}. Other models, such as those 
dominated by large cold dark matter haloes containing many galaxies
could also in principle compute their resulting galaxy peculiar
velocity distribution functions and compare them with the observed
distribution function, but they do not yet appear to have examined
this question in detail.

This distribution function in velocity (or momentum) space complements
the distribution in configuration space, and their consistent
combination provides a statistical description of evolution in the
complete six-dimensional phase space. Originally, \citep{saslaw10},
the velocity distribution was derived from the spatial distribution by
making the additional main assumption that on average the potential
energy fluctuations are proportional to the kinetic energy
fluctuations in any local volume. Direct $N$-body simulations
confirmed this as a good approximation.  Recently a new and more
general statistical mechanical derivation \citep{ahmad} of the
spatial distribution function has been calculated. In the present
paper, we show how the earlier assumption relating fluctuations can
now be proven directly from the statistical mechanical partition
function, thus deriving the velocity distribution more
rigorously. Moreover, the statistical mechanical derivation contains
an explicit softening parameter for the gravitational potential. This
provides a simple ``isothermal sphere'' model for individual galaxy haloes and
allows us to determine the effects of such haloes on the peculiar
velocity distribution. Comparison with observations gives an
approximate upper limit to the size of such haloes. 

Combining the velocity distribution function with the new statistical
mechanical approach also leads to relatively simple solutions of some other
fundamental questions. The additional questions we will examine here
are: What is the probability that $N$ galaxies in a randomly placed
volume of size $V$ form a bound cluster? What is the probability that
such a bound cluster is virialized? What is the probability of forming
a small group of galaxies around which the peculiar velocity
dispersion is so small that the local flow agrees well
with the global Hubble expansion, as is observed around our local
cluster?

To answer these questions, we calculate the density of energy states
for the gravitationally interacting cosmological many body
system. This is one of a very few statistical mechanical  interacting
systems (the Ising model is another) for which such a calculation has
been possible. 

In \S 2, we derive what was previously the basic assumption that local
kinetic energy fluctuations are proportional to local potential energy
fluctuation. We then obtain the velocity distribution function for the softened
potential representing  simple isothermal haloes around individual
galaxies. Comparison with the observed peculiar radial velocity
distribution function provides an estimate for the maximum average
size of such haloes. \S 3 calculates the density of energy states and
the probabilities for bound and virialized clusters, and \S 4
discusses implications for the cool local Hubble flow. Finally, \S 5
briefly summarizes and discusses our results. 

%%%%%%%%%%%%%%%%%%%%%%%%%%%%%%%%%%%%%%%%%%%%%%%%%%%%%%%%%%%%%%%%%
\section{DERIVATION OF THE VELOCITY DISTRIBUTION FUNCTION INCLUDING
GALAXY HALOES}  \label{sec2}
%%%%%%%%%%%%%%%%%%%%%%%%%%%%%%%%%%%%%%%%%%%%%%%%%%%%%%%%%%%%%%%%%%

To derive the velocity distribution, we start with the canonical partition
function derived for the cosmological many body problem viewed as a
model for galaxy clustering \citep{ahmad, saslaw2003}:
\begin{equation} \label{e:ZN1}
\begin{split}
Z_{N} (T,V) 
&= \frac{1}{\Lambda^{3N} N!} \int \exp \l\{-\l[\sum^{N}_{i=1}
\frac{p_i^2}{2m} + \phi (r_1, r_2, \dots r_N) \r] T^{-1} \r\} d^{3N} p~
d^{3N} r \\
&= \frac{1}{N!} \l(\frac{2\pi m T}{\Lambda^2}
\r)^{\frac{3N}{2}} V^{N} \l[ 1 + \kappa \bar{n} T^{-3} \zeta
\l(\frac{\epsilon}{R_1} \r) \r]^{N-1} ~, 
\end{split}
\end{equation}    
where $\Lambda$ is a normalization factor, to be discussed in \S 3, 
\begin{equation} \label{e:ZN2}
\kappa=3/2(Gm^2)^3~,
\end{equation}
assuming all $N = \bar{n} V$ galaxies in the volume $V$ have an
average mass $m$ and temperature $T$, and 
\begin{equation} \label{e:zeta}
\zeta \l(\frac{\epsilon}{R_1} \r) = \sqrt{1+ \l(\frac{\epsilon}{R_1}
\r)^2} + \l(\frac{\epsilon}{R_1} \r)^2 \ln
\frac{\frac{\epsilon}{R_1}}{1+ \sqrt{1+ \l(\frac{\epsilon}{R_1}\r)^2}} ~.
\end{equation}
Here $\epsilon$ is the softening parameter for the
gravitational potential energy
\begin{equation} \label{e:phiij0}
\phi (r_1, r_2, \dots r_N) = \sum_{1 \leq i \leq j \leq N} \phi
(r_{ij}) ~,
\end{equation}
with
\begin{equation} \label{e:phiij}
\phi_{ij} = - \frac{G m^2}{\sqrt{r_{ij}^2 + \epsilon^2}} ~,
\end{equation} 
and $R_1$ is a comparison scale which may, for example, be the average
distance between those galaxies which contribute most of the mass. 

When $r_{ij} < \epsilon$, the potential energy is approximately constant, as
would characterize an approximately isothermal sphere with density
$\rho \propto r^{-2}$. Of course, this can be generalized easily with
a range of values of $\epsilon$ for galaxies of different masses, but here
we are interested in basic results which can be obtained more simply
by considering all the galaxies to have their average values of
$\epsilon$ and $m$. 

To examine fluctuations in the numbers and energies of galaxies
which can move between cells, we need the grand canonical ensemble
whose partition function is
\begin{equation} \label{e:ZG1}
Z_{G} (T,V) = \sum^{\infty}_{N=0} e^{\frac{N \mu}{T}} Z_{N} (T,V) ~,
\end{equation}
where $\mu$ is the chemical potential derived from $Z_N$. The grand
canonical partition function is
related to the equation of state by
\begin{equation} \label{e:ZG2}
\ln Z_G = \frac{PV}{T} = \bar{N} (1 - b_{\epsilon}) ~,
\end{equation}
where
\begin{equation} \label{e:bepsilon}
b_{\epsilon} = \frac{b \zeta \l(\frac{\epsilon}{R_1} \r)}{1+ b
\l[\zeta \l(\frac{\epsilon}{R_1} \r) -1\r]} 
=  \frac{\kappa \bar{n} T^{-3}
\zeta \l(\frac{\epsilon}{R_1} \r) }{ 1 + \kappa \bar{n} T^{-3} \zeta
\l(\frac{\epsilon}{R_1} \r)  } ~,
\end{equation}
with, as usual,
\begin{equation} \label{e:bparameter}
b = - \frac{W}{2K} = \frac{2 \pi Gm^2 \bar{n}}{3T} \int^{R}_{0} \xi_2
(r) r dr 
\end{equation}
the ratio of gravitational correlation energy, $W$, to twice
the kinetic energy, $K$, of peculiar velocities in a volume, here
taken to be spherical with radius $R$. The two galaxy correlation
function is $\xi_2 (r)$. 

From equations $\eqref{e:ZN1}$-$\eqref{e:bparameter}$, the spatial
distribution function, which is the sum over all energy states in the
grand canonical ensemble becomes \citep{ahmad}:
\begin{equation} \label{e:GQED1}
\begin{split}
f(N, b_{\epsilon}) &= \frac{1}{Z_{G}} \l(\sum^{N}_{i=0} e^{\frac{N \mu}{T}}
e^{-\frac{E_i}{T}} \r) \\
&= \frac{\bar{N} (1- b_{\epsilon}) }{N!} [\bar{N} (1-
b_{\epsilon}) + N b_{\epsilon} ]^{N-1} e^{-[\bar{N} (1-
b_{\epsilon}) + N b_{\epsilon}]} ~,
\end{split}
\end{equation}
where the chemical potential is given by
\begin{equation} \label{e:chemical}
\begin{split}
\frac{\mu}{T} 
&= \frac{1}{T} \l( \pp{F}{N} \r)_{T,V} \\
&= \ln \l(\frac{\bar{N}}{V} T^{-\frac{3}{2}} \r) + \ln
(1-b_{\epsilon}) - b_{\epsilon} - \frac{3}{2} \ln \l(\frac{2 \pi
m}{\Lambda^2} \r) ~. 
\end{split}
\end{equation}
In the limit $\epsilon \to 0$ so that $\zeta \to 1$ and $b_{\epsilon}
\to b$, $f(N, b_{\epsilon})$ becomes
the distribution function for point masses \citep{saslaw2}.

To derive the gravitational velocity distribution function, $f(v)$, from the spatial
distribution, $f(N, b_{\epsilon})$, we need to modify the grand canonical
ensemble by adding a relation between $N$ and $v$. Previously in \cite{saslaw10}, 
this was postulated by assuming that over a given volume any local fluctuation 
of kinetic energy (caused by correlations among particles) is proportional to 
its potential energy fluctuation giving
\begin{equation} \label{e:velocity1} 
N \l\langle \frac{1}{r_{ij} } \r\rangle_{\text{Poisson}} = \alpha v^2 ~.
\end{equation}
The detailed configuration in each fluctuation is represented by the
form factor $\alpha$, indicating the local departure from
\begin{equation} \label{e:velocity2}
\l\langle \frac{1}{r_{ij}} \r\rangle_{\text{Poisson}} = \l(\frac{4 \pi}{3}
n \r)^{\frac{1}{3}} \Gamma\l( \frac{2}{3} \r) = 2.18 n^{\frac{1}{3}}
\end{equation}
for a Poisson distribution (see Saslaw 1987, equation 33.23 with
$a=0$). Although $\alpha$ varies among fluctuations, it clusters
strongly around its average value, as n-body simulations show
\citep{saslaw10}. 

Instead of postulating $\eqref{e:velocity1}$, we can now derive it from
the configuration integral in $\eqref{e:ZN1}$:
\begin{equation} \label{e:configuration}
\begin{split}
Q_N (T, V) &\equiv \int \dots \int \exp [-\phi(r_1, r_2, \dots, r_N) T^{-1}]
d^{3N} r \\
&= V^{N} \l[ 1 + \kappa \bar{n} T^{-3} \zeta \l(\frac{\epsilon}{R_1}
\r) \r]^{N-1} ~.
\end{split}
\end{equation}
The local fluctuation of the potential energy around the spatially
constant mean field in a volume containing $N$ galaxies therefore has
an ensemble average value:
\begin{equation} \label{e:phi}
\begin{split} 
\langle \phi \rangle &= -\frac{1}{Q_N} \pp{Q_N}{T^{-1}} 
= - \pp{ (\ln Q_n)}{T^{-1}}  \\
&= -3(N-1) b_{\epsilon} T  \\
&= - (N - 1) b_{\epsilon} m \langle v^2 \rangle
\end{split}
\end{equation}
using $\eqref{e:configuration}$ and $\eqref{e:bepsilon}$. We may also
write the average potential energy in this volume as 
\begin{equation} \label{e:seperation1}
\langle \phi \rangle = - \sum_{1 \leq i < j \leq N} \frac{G
m^2}{r_{ij}} = - \frac{G m^2 N (N-1)}{2} \l\langle \frac{1}{r_{ij}}
\r\rangle  ~,
\end{equation}
where 
\begin{equation} \label{e:seperation2}
 \l\langle \frac{1}{r_{ij}} \r\rangle = \frac{2}{N(N-1)}   \sum_{1
 \leq i < j \leq N} \frac{1}{r_{ij}} = \eta  \l\langle \frac{1}{r_{ij}}
\r\rangle_{\text{Poisson}}   ~.
\end{equation}
Here $\eta$ represents the form factor. From
$\eqref{e:phi}$-$\eqref{e:seperation2}$, we obtain
\begin{equation} \label{e:seperation3}
N = \frac{2 b_{\epsilon}}{Gm \eta} \l\langle \frac{1}{r_{ij}}
\r\rangle_{\text{Poisson}}^{-1} \langle v^2 \rangle  ~,
\end{equation}
which is the same as $\eqref{e:velocity1}$ averaged over many cells having
the same $N$ and $\alpha$ with
\begin{equation} \label{e:alpha}
\alpha = \frac{2 b_{\epsilon}}{G m \eta} ~.
\end{equation}

In regions of enhanced density, the $r_{ij}$ are generally smaller
than their Poisson values for the average density $\bar{n}$ and
consequently $\eta \gtrsim 1$ from $\eqref{e:seperation2}$. Using
natural units $G=m=R=1$, equation $\eqref{e:alpha}$ shows that for the
currently observed clustering with $b \approx 0.75$, the value of
$\alpha$ is also of order unity. In these units, averaging
$\eqref{e:seperation3}$ over the entire system for individual galaxies
so that $\langle N \rangle = 1$ gives the approximate relation with
$\beta = \langle v^2 \rangle$:
\begin{equation} \label{e:alphafactor}
\alpha \approx  \l\langle \frac{1}{r_{ij}}
\r\rangle_{\text{Poisson}} \beta^{-1} ~.
\end{equation}
Although this average value of $\alpha$ (or equivalently $\eta$)
simplifies the description, it gives a reasonably accurate result since the
velocity distribution involves all the galaxies in the system, and is
thus an effective average over all the cluster configurations in the
entire ensemble. 

\begin{center}
\begin{figure}[!bth]
\plotone{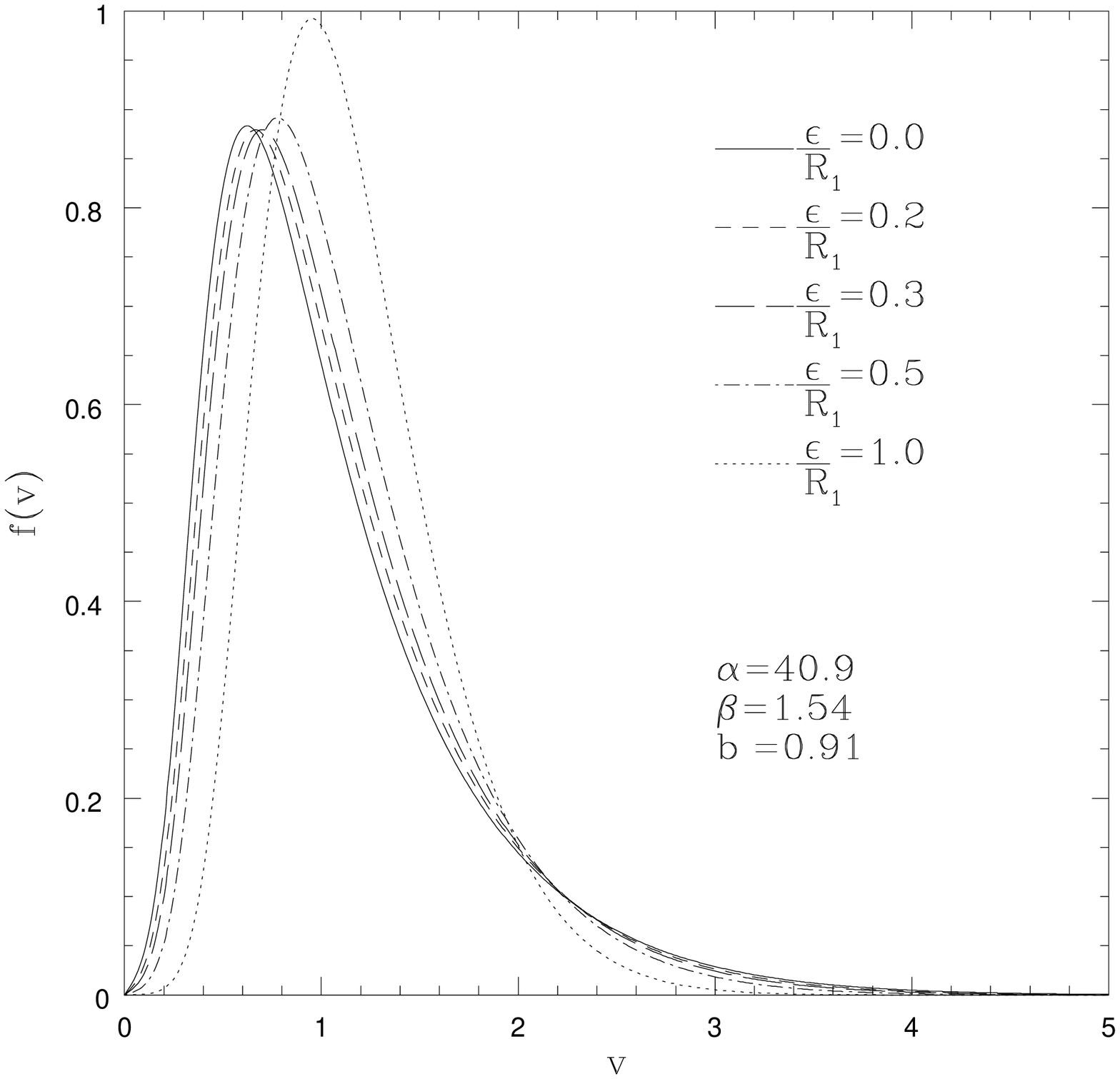}
\caption{Peculiar velocity distribution functions using observed
values of $\alpha$, $\beta$ and b (Raychaudhury and Saslaw 1996)
from the sample of 825 Sb-Sd galaxies with distance $D<5000$ km s$^{-1}$ from
the Mathewson et al catalogue with peculiar velocities corrected for
the Dressler et al. (1986) bulk motion of 599 km s$^{-1}$. The
velocity distributions of equation $\eqref{e:GQEDV1}$ are compared for
a range of values of the halo size, $\epsilon/R_1$.} 
\label{veldistb}
\end{figure}
\end{center}

To transform the spatial distribution $f(N)$ in $\eqref{e:GQED1}$ into
the velocity distribution function, we follow the same procedure as for
$\epsilon=0$ \citep{saslaw10} to obtain
\begin{equation} \label{e:GQEDV1}
f(v) dv = \frac{2 \alpha^2 \beta (1-b_{\epsilon})}{\Gamma(\alpha v^2
+1)} [\alpha \beta (1-b_{\epsilon}) + \alpha b_{\epsilon}
v^2]^{\alpha v^2 -1} e^{-[\alpha \beta (1-b_{\epsilon}) + \alpha b_{\epsilon}
v^2]} v dv ~.
\end{equation}
Here $\Gamma (\alpha v^2 +1)$ is the usual gamma function. Figure \ref{veldistb}
shows examples of $f(v)$ for some typical parameters suggested by the
observations and for a range of $\epsilon$. For $\epsilon=0$, we
obtain the previous results. As $\epsilon$ increases, equations
$\eqref{e:zeta}$ and $\eqref{e:bepsilon}$ show that $\zeta
(\epsilon/R_1)$ decreases and that $b_{\epsilon}$ increases
monotonically with increasing $b$ and also with increasing $\zeta
(\epsilon/R_1)$. Moreover, $b > b_{\epsilon}$, and the disparity
between them will increase as $\epsilon$ increases, though for
$\epsilon/R_1 < 0.5$, the value of  $b - b_{\epsilon} < 0.1$. As
$\epsilon$ increases, Fig. \ref{veldistb} shows that the peak of
$f(v)$ shifts towards higher velocities, but there are fewer galaxies
in both the low and high velocity tails, relative to the point mass
case $\epsilon=0$. This effect only becomes substantial for $\epsilon
\gtrsim 0.5$, when the haloes of individual galaxies overlap with the
haloes of their nearest neighbours. 

The effects of halo overlap are to smooth out the gravitational
potential wells so that close encounters produce fewer high velocity
galaxies. The reason that there are also fewer low velocity galaxies
as $\epsilon$ increases is that the forces are more uniformly
distributed throughout the system and fewer galaxies are so
gravitationally isolated that their peculiar velocities simply decay
adiabatically as the universe expands. 

These effects are also illustrated nicely by the radial velocity
distribution function, which is what we usually observe. The total
velocity $v$ is related to its radial and tangential components by 
\begin{equation} \label{e:defv}
v^2 = v_r^2 + v_{\theta}^2 + v_{\phi}^2 \equiv v_r^2 + v_{\bot}^2 ~.
\end{equation} 
Integrating $\eqref{e:GQEDV1}$ over the tangential velocities
\citep{inagaki} now gives the radial velocity distribution function for
the softened potential:
\begin{equation} \label{e:GQEDV2}
\begin{split}
f(v_r) &= \alpha^2 \beta (1- b_{\epsilon}) e^{-\alpha \beta
(1-b_{\epsilon})} \\
&\times \int^{\infty}_{0} \frac{v_{\perp}}{\sqrt{v_r^2 +
v_{\perp}^2}} \frac{[\alpha \beta(1- b_{\epsilon})+ \alpha
b_{\epsilon} (v_r^2 + v_{\perp}^2)]^{\alpha(v_r^2 + v_{\perp}^2)-1}
}{\Gamma[\alpha(v_r^2 + v_{\perp}^2) +1]]}  e^{-\alpha
b_{\epsilon} (v_r^2 + v_{\perp}^2)} d 
v_{\perp} 
\end{split} 
\end{equation}
where again $\beta \equiv \langle v_r^2 + v_{\bot}^2 \rangle$.  

\begin{center}
\begin{figure}[!bth]
\plotone{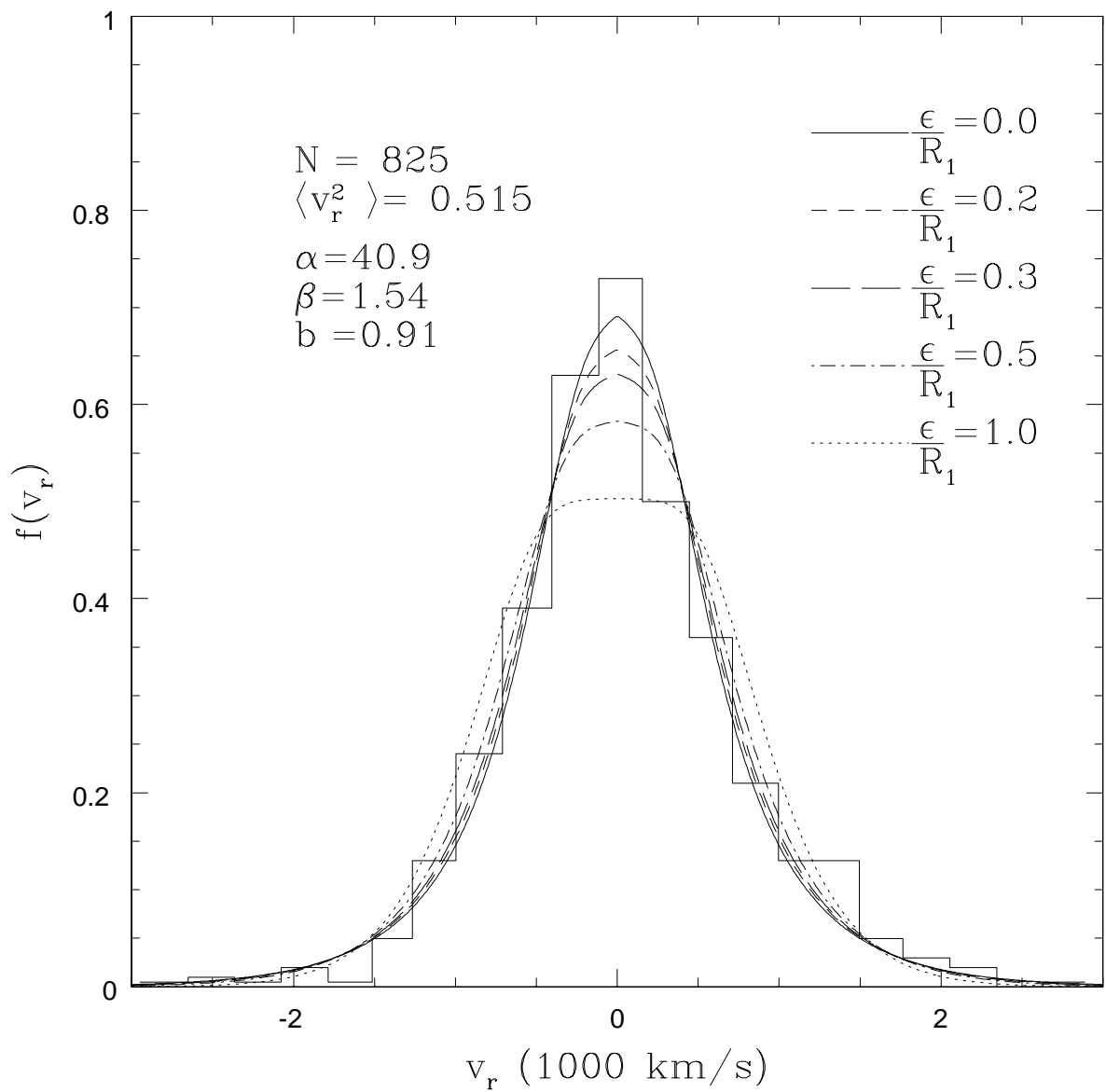}
\caption{The observed distribution (histogram, Raychaudhury and Saslaw
1996) of radial peculiar velocities for the sample of 825
Sb-Sd galaxies with distance $D<5000$ km s$^{-1}$ from the Mathewson
et al catalogue, with  peculiar velocities corrected for the
Dressler et al. (1986) bulk motion of 599 km s$^{-1}$. The radial
velocity distributions of equation $\eqref{e:GQEDV2}$ are compared for a range of
values of the halo size, $\epsilon/R_1$.} 
\label{rpvx1b}
\end{figure}
\end{center}

Fig. \ref{rpvx1b} shows the observed histogram of radial peculiar
velocities for a sample of 825 Sb-Sd galaxies with $v_{\text{radial}}
\leq 5000$ km s$^{-1}$ from Figure 2a in
\cite{raychaudhury}, as an illustration. The solid line is the best
fit to this histogram for $\epsilon=0$; it has $\alpha=40.9$,
$\beta=1.54 \approx 3 \langle v_r^2 \rangle$, and $b=0.91$. The broken
lines show the effect of increasing $\epsilon/R_1$ while holding the
other parameters constant. 

Evidently only relatively large haloes, $\epsilon/R_1 \gtrsim 0.3$,
affect the velocity distribution (or the counts in cells)
significantly. Since these large haloes soften and reduce the
gravitational forces, they lead to reduced clustering. This implies
fewer galaxies in the very high (radial) velocity tail of the
distribution and a broadening around the most probable velocity (at
zero for $v_r$) whose peak is shifted to somewhat higher
velocities. This leaves fewer underdense regions with low
velocities. If there are fewer rich clusters and the number of
galaxies is at least approximately conserved (i.e. mergers do not
dominate, of if they do , they mostly occur near the centers of mass
of the merging galaxies and these centers have approximately the same
distribution as the galaxies themselves) then fewer voids and
underdense regions are formed to create the clusters.   

As an illustration, if we take $R_1$ to be an average distance of
$\sim 1$ Mpc between typical massive galaxies, this would suggest that
haloes would need to be about 300 kpc in radius for a significant
effect. These numbers are made uncertain by the criterion adopted for
massive galaxies and the lack of precision in determining the
``effective radius'' of a halo whose density falls off as $r^{-2}$,
but they are indicative. If $\epsilon/R_1 \approx 0.5$, the haloes are
nearly touching, and if $\epsilon/R_1 \gtrsim 1$, several galaxies
would share a common halo. The latter case, although it is favoured by
some cold dark matter scenarios, would not appear to agree with the
presently observed velocity distribution. Instead of the orbits being
mainly determined by the (softened) gravitational forces of individual
galaxies, such models with $\epsilon/R_1 \gtrsim 1$ determine the
orbits mainly by the inhomogeneous distribution of the matter in the
common halo. For a communal halo which was not highly concentrated,
this would lead to a radial velocity distribution which is relatively
broader and less peaked compared to the case with individual haloes and
small $\epsilon/R_1$. 

In Fig. \ref{rpvx1b}, by setting $\epsilon/R_1=0$, we determine the
best fit values of $b, \alpha$ and $\beta$. Including effects of dark
matter haloes, observations of the spatial distribution function,
equation $\eqref{e:GQED1}$, or the variance of its counts in cells, gives 
$b_{\epsilon}$ directly. Observations of $\xi_2 (r)$, $\bar{n}$, the
peculiar velocity dispersion and an estimated average galaxy
mass $m$ (including haloes) would provide $b$ from
equation $\eqref{e:bparameter}$. Then equations $\eqref{e:seperation2}$ and
$\eqref{e:alphafactor}$ give $\alpha$ and all the quantities in the
theory can, in principle, be found consistently from observations. Comparing
$b_{\epsilon}$ and $b$ yields $\epsilon/R_1$.  

Figure 2 in \cite{ahmad} shows that for $\epsilon/R_1 \lesssim 0.5$,
the values of $b$ and $b_{\epsilon}$ differ by less than
0.07. Although this difference is small, our Figure 2 shows it can
significantly affect the peak of the radial velocity distribution. In
practice there is not yet enough information to determine
$\epsilon/R_1$ accurately, and it can only be fitted within ranges for
prescribed confidence limits. Current data, consistent with
$\epsilon/R_1 \lesssim 0.5$, suggest that most individual galaxies are
surrounded by their own haloes rather than embedded in large massive
common haloes.  

\section{PROBABILITIES FOR BOUND AND VIRIALIZED CLUSTERS}

\subsection{The Density of Energy States}

For estimates of the mass-luminosity ratio or the amount of dark
matter in clusters of galaxies, it is generally assumed that the
clusters are gravitationally bound and have a high probability of
being virialized \citep{saslaw3, saslaw2}. Our
statistical mechanical analysis now makes it possible to calculate
these probabilities directly for the cosmological many-body problem,
and to determine how they are affected by the value of $b$ and the
average peculiar velocity dispersion (or temperature), as well as by
the average halo size $\epsilon/R_1$. 

The general probability density, in a grand canonical ensemble, for a region
with $N$ galaxies to be in the differential range of energy states,
$dE$, follows from the first equality of equation $\eqref{e:GQED1}$
without the summation over energies as 
\begin{equation} \label{e:prob1}
P(E, N, T_{0}, V, \mu) dE = g(E) \frac{e^{-E/T_{0}} e^{N\mu/T_{0}} }{Z_G
(T_{0},V,\mu)} dE
\end{equation}
where $g(E)$ is the density of energy states and $Z_G$ is given by
equation $\eqref{e:ZG2}$. The continuous density of states arises from
transforming the sum over individual energy microstates $E_i$ into an
integral over a finite range of energies. Here $T_{0}$ is the average
temperature for the ensemble.

To determine $g(E)$ for cells containing $N$ galaxies, we first
recall, for example from \cite{greiner}, that a subset of the grand
canonical ensemble in which each 
cell contains exactly $N$ galaxies (particles) forms a canonical
ensemble. It is in contact with a temperature reservoir so that
energy, but not particles, can move among cells. A particular cell of
volume $V$, with a particular configuration of $N$ galaxies has an
energy $E$. Averaging these energies over all cells gives the average
energy $U$ which appears in the equation of state for the system 
\citep{saslaw9, ahmad}
\begin{equation} \label{e:internalenergy3}
U = \frac{3}{2} NT (1- 2 b_{\epsilon})~,
\end{equation}
since if all cells have a  given value of $N$, the grand canonical
equation of state also applies to this canonical ensemble. Boltzmann's
fundamental postulate of statistical mechanics relates $g(E)$ to the
entropy in the microcanonical ensemble by 
\begin{equation} \label{e:omegaE}
\Omega(E) = \int_{\Delta E} g(E) dE = e^{S(E, V, N)}
\end{equation}
where $\Omega(E)$ is the number of energy states in a small range
$\Delta E \ll E$, so $\Omega(E) = g(E) \Delta E$. 

It is well known that for large values of $N$, the energy $E$ in the
microcanonical ensemble is very nearly equal to the average energy $U$
in the canonical ensemble. Under this condition, we can examine
canonical ensembles with different average energies $U$ to determine
$P(E)$ from equation $\eqref{e:prob1}$.  Therefore we next determine
the detailed condition for $S(E, V, N) \approx S(U, V, N)$ to hold. 

The energy dependence in equation
$\eqref{e:prob1}$, which involves both the density of states and the
Boltzmann factor, is the free energy $-F = T~S(E) - E$. The
probability of any energy will therefore have an extremum at 
\begin{equation} \label{e:condition3}
T \l(\pp{S}{E} \r)_{E=U} = 1~.
\end{equation}
This extremum will be a maximum if 
\begin{equation} \label{e:extremum}
\l(\ppd{S}{E} \r)_{E=U} = \l(\pp{}{E} \frac{1}{T} \r)_{E=U} =
-\frac{1}{T^2} \l( \pp{T}{E} \r)_{E=U} = - \frac{1}{NT^2 C_V} < 0
\end{equation}
where $C_V$ is the specific heat of the entire ensemble. This ensemble will
contain a range of cluster masses and energies for any particular value of $b_{\epsilon}$. 
As equation $\eqref{e:fluctuationterm2}$ indicates, 
$C_V$ will remain positive for $b_{\epsilon}$ less than 0.86.  If $C_V$ becomes
negative, $P(E)$ will have no maximum. This state represents a system dominated by negative energy
clusters which are however virialized so that they do not become unstable on short
timescales and the system remains in quasi-equilibrium. When there is a maximally probable
state for energy $E$, we expand the free energy around it:
\begin{equation}
\begin{split}\label{e:fluctuationterm}
S(E) - \frac{E}{T} &= \l[S(U) - \frac{U}{T} \r] + \frac{1}{2} (E-U)^2
\l(\ppd{S}{E} \r)_{E=U} + \dots \\
&= \l[S(U) - \frac{U}{T}\r] - \frac{1}{2 NT^2 C_V} (E-U)^2 + \dots 
\end{split}
\end{equation}
Hence the departures of $S(E)$ and $E$ from $S(U)$ and $U$ are closely
related to the variance of the energy fluctuations around the
mean. Using earlier results for $C_V$ and $\langle (\Delta E)^2
\rangle$ in the grand canonical ensemble \citep{saslaw9, ahmad}, which
give an upper limit to the fluctuation term in equation
$\eqref{e:fluctuationterm}$, we find
\begin{equation} \label{e:fluctuationterm2}
\frac{1}{2 NT^2 C_V} (E-U)^2 = \frac{1}{4(1-b_{\epsilon})^2} 
\frac{(5-20 b_{\epsilon} + 34 b_{\epsilon}^2 - 16
b_{\epsilon}^3)}{(1+4 b_{\epsilon} - 6 b_{\epsilon}^2)} ~.
\end{equation}  
This ratio has two poles at $b_{\epsilon} = (2 + \sqrt{10})/6 = 0.86$
when $C_V$ passes through zero and at $b_{\epsilon} = 1$ when all the
galaxies collect into one nearly virialized cluster
\citep{baumann}. Away from these poles, the fluctuation term decreases
rapidly (e.g. it is 44.5 at $b_{\epsilon} = 0.8$, 15.2 at
$b_{\epsilon} = 0.75$ and 3.78 at $b_{\epsilon} = 0.5$) compared to
$S(U) - U/T$ which is of order $N [1+3b + \ln (1-b) +$ non-gravitational
terms$]$. Since groups must contain $N \geq 2$ galaxies, the
approximation $S(E) \approx S(U)$ is usually reasonable. This is
closely related to the neglect of terms of order $\ln N$ in the
thermodynamic limit of large $N$ since equation $\eqref{e:omegaE}$
then gives $S = \ln g(E) + \ln \Delta E \approx \ln g(E)$. Taking
$\Delta E$ to be a unit dimensionless variation of the energy shell,
$\Delta E_{*}$, from equation $\eqref{e:Estar}$ below gives $S = \ln
g(E_{*})$. 

The grand canonical entropy, which for a sub-ensemble of fixed $N$ is
the same as the canonical entropy, is a function of temperature
\citep{saslaw9, ahmad}: 
\begin{equation} \label{e:entropy}
S(T, V, N) = -N \ln \l(\frac{N}{V T^{\frac{3}{2}}} \r) + N \ln [1 +
aT^{-3}] - \frac{3N a T^{-3}}{1 + a T^{-3}} + \frac{5}{2}N +
\frac{3}{2} N \ln \l(\frac{2 \pi m}{\Lambda^2} \r)
\end{equation} 
where from equations $\eqref{e:ZN2}$ and $\eqref{e:bparameter}$
\begin{equation} \label{e:a1}
a \equiv \frac{3}{2} (Gm^2)^3 \bar{n} \zeta \l(\frac{\epsilon}{R_1}\r)   
\end{equation}
has dimensions of (energy)$^3$. In fact $a^{1/3}$ is essentially the
absolute value of the potential energy of two galaxies with average
seperation. The relation between $T$ and $E$ for $U \to E$ in volumes
with energy $E$ and given $N$ is given by the energy equation of state
$\eqref{e:internalenergy3}$. Normalizing to dimensionless variables, 
\begin{equation} \label{e:Estar}
E_{*} \equiv \frac{2 U}{3 N a^{\frac{1}{3}} } ~,
\end{equation} 
and 
\begin{equation} \label{e:Tstar}
T_{*} \equiv \frac{T}{a^{\frac{1}{3}}} ~.
\end{equation}
Here $T_{*}$ is defined as the local fluctuating temperature within
the ensemble, which differs from
$T_{0*} \equiv T_{0}/a^{1/3}$, the average dimensionless temperature
of the whole ensemble. Using equation $\eqref{e:bepsilon}$, equation
$\eqref{e:internalenergy3}$ becomes 
\begin{equation} \label{e:Estareqn}
E_{*} = \frac{T_{*} (T_{*}^3 - 1) }{T_*^3 + 1} ~,
\end{equation}
or 
\begin{equation} \label{e:Tstareqn}
T_{*}^4 - E_{*} T_{*}^3 - T_{*} - E_{*} = 0 ~.
\end{equation}
The solutions of both the linear equation for $E_{*}(T_{*})$ and the
quartic for $T_{*} (E_{*})$, although equivalent, provide different
insights into the relation between $T$ and $E$. 

Since equation $\eqref{e:Estareqn}$ is simpler, we consider it
first. Figure 3 shows $E_{*}(T_{*})$. It is double valued between the
zeroes at $T_{*} =0$ and $T_{*} = 1$ and has a minimum at 
\begin{equation}
T_{*} = \l(\frac{2}{3} \sqrt{3} -1\r)^{\frac{1}{3}} = 0.54,
\quad E_{*} = -0.390 ~.
\end{equation}
This value of  $T_{*}$ will be found to divide the two real
branches of the solutions to the quartic equation
$\eqref{e:Tstareqn}$. For large values of $T_{*}$, produced either by
a high temperature or a weak interaction $a^{1/3}\to 0$, $E_{*} =
T_{*}$ in equation $\eqref{e:Tstareqn}$, which becomes the perfect gas
equation of state. For small $T_{*}$, the limit of equation
$\eqref{e:Tstareqn}$ is $E_{*} = - T_{*}$ which is the equation of
state for a completely virialized system having specific heat $C_V =
-3/2$. 

The real solutions of the quartic equation $\eqref{e:Tstareqn}$ are 
\begin{equation} \label{e:Tsolution}
T_{* \pm} (E_{*}) = \frac{E_{*}}{4} + \frac{1}{2}
\sqrt{\frac{E_{*}^2}{4} + Z_1} \pm \frac{1}{2} \sqrt{q} 
\end{equation}  
where 
\begin{equation} \label{e:zfn1}
Z_1 = 
\l(\frac{1}{2} - \frac{E_{*}^3}{2} - \sqrt{\frac{1}{4} + \frac{223
E_{*}^3}{54} + \frac{E_{*}^6}{4} } \r)^{\frac{1}{3}}
+ \l(\frac{1}{2} - \frac{E_{*}^3}{2} + \sqrt{\frac{1}{4} + \frac{223
E_{*}^3}{54} + \frac{E_{*}^6}{4}} \r)^{\frac{1}{3}}
\end{equation}
and 
\begin{equation} \label{e:q1}
q =  \l(\frac{E_{*}}{2} + \sqrt{\frac{E_{*}^2}{4} + 
Z_1} \r)^2 - 2 Z_1 + 2 \sqrt{Z_1^2 + 4 E_{*}} 
\end{equation}

The number of energy states from equations $\eqref{e:omegaE}$ and
$\eqref{e:entropy}$ is 
\begin{equation} \label{e:omegaE2}
\Omega(T_{*}) = \l(\frac{2 \pi m a^{\frac{1}{3}} }{\Lambda^2}
\r)^{\frac{3N}{2}} \l(\frac{V}{N} \r)^N e^{5N/2}~T_{*}^{3N/2}~[1 +
T_{*}^{-3}]^N~e^{- 3N/(1+ T_*^3)}~.
\end{equation}
We then find $T_{*} (E_{*})$ from either equations $\eqref{e:Estareqn}$
or $\eqref{e:Tsolution}$-$\eqref{e:q1}$ to calculate $g(E_*) = d
\Omega(E_*) / dE_{*} \approx \Omega(E_*)/\Delta E_*$. In fact, we use
the results from both these solutions to double check the numerical
computations; they agree with each other. Already from equation
$\eqref{e:omegaE2}$, we see for $T_{*}^3 \gg 1$ and thus $T_* \approx
E_*$ that $g(E_*) \propto E_*^{(3N/2)- 1}$ which agrees with the
usual energy dependence of $g(E)$ in an ideal gas \citep{greiner}. For
the dual limit as $T_* \to 0$ in equation $\eqref{e:Tstareqn}$ and $T_{*} \to -
E_{*}$, recalling that the potential energy dominates so that $E_{*} < 0$,
we obtain the same energy dependence of $g(E_*)$. This describes a
dense or massive system in which $a^{1/3} \to \infty$, even though T may
remain large and virialized, and the system behaves approximately as
an idealized gas gravitationally bound in a cluster. A system of
approximately isothermal spheres each with the same nearly
Maxwell-Boltzmann velocity distribution would be an example. In
equation $\eqref{e:omegaE2}$, we may normalize the volume $V$ and the
mass $m$ to unity by selecting the phase space cell normalization
$\Lambda^2$, as often done for the classical gas \citep{sommerfeld}.

\begin{center}
\begin{figure}[!bth]
\plotone{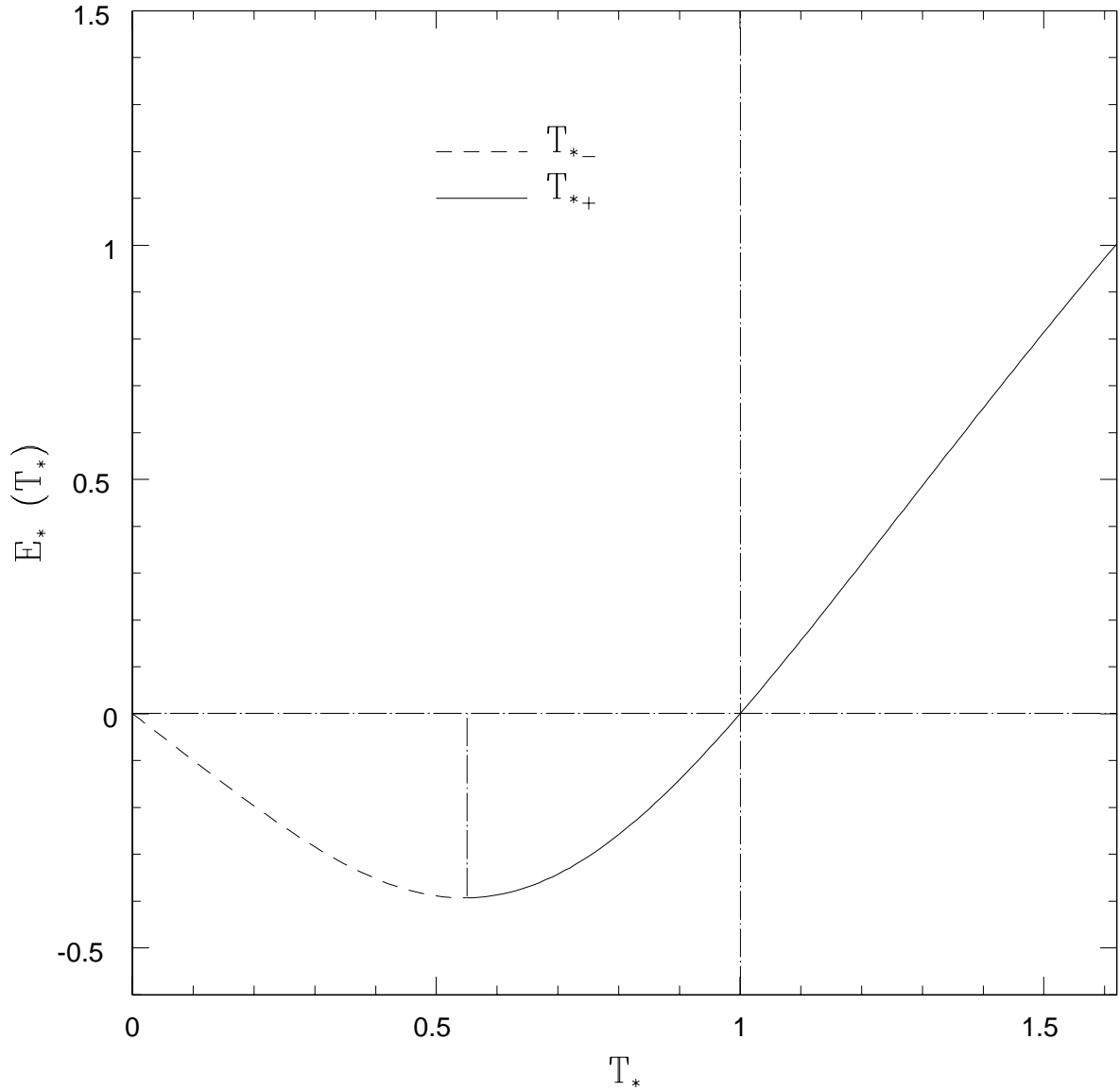}
\caption{The relation between $T_{*}$ and $E_{*}$. The
curve corresponds to the two real branches of the four solutions in the
quartic equation $\eqref{e:Tstareqn}$. The curve reaches a minimum at
$T_{*}=0.54$ and $E_{*}=-0.39$. The region $T_{*} \leq
1.0$ represents bound groups. The solid line represents the $T_{*+}$
branch and the dashed line represents the $T_{*-}$ branch having real
valued energy.}  
\label{xt}
\end{figure}
\end{center}

Generally we solve for $g(E_*)$ numerically from $\eqref{e:omegaE2}$
and either $\eqref{e:Estareqn}$ or $\eqref{e:Tstareqn}$ and then
integrate equation $\eqref{e:prob1}$ over a range of energies (or equivalent temperatures) of
interest. The normalization is such that for fixed $N$, $T_{0*}$, $V$ and $\mu$ 
\begin{equation} \label{e:norm}
\int^{E_{* \text{max}}}_{E_{* \text{min}}} P(E_*, N, T_{0*}, V, \mu) d
E_* = 1~. 
\end{equation}
It is important to keep in mind, because $T_{*} (E_*)$ has a double valued
regime as in Figure \ref{xt}, that the subscripts "max" and "min" refer
to the maximum and minimum values of $T_{*}$ over which the integral is
evaluated, and not to the maximum and minimum values of the energy. Thus the
integral is evaluated from a value of $E_{* ~\text{min}}= E_{*} (T_{*
~\text{min}})$ corresponding to the minimum value on the $T_{* -}$ branch,
through $E_{*~\text{virial}}=-0.33$ on the $T_{* +}$ branch, and then to
$E_{* \text{max}} = E_{*} (T_{* ~\text{max}})$ after switching to the $T_{*+}$
branch at $T_{*} =0.54$ and $E_{*} = -0.390$. 

Since the system is in quasi-equilibrium in the expanding universe,
its range of energies, $E_{* \text{min}} \leq E_* \leq E_{*
\text{max}}$, is more restricted than the usual infinite range. States
with positive energies represent unbound groups or clusters, and if
their energy is too great they will fly apart on time-scales much
shorter than the Hubble time and be inaccessible to the thermodynamics
of the quasi-equilibrium system. Negative energy states which are not
near virial equilibrium will collapse on time-scales much shorter than
the Hubble time, and they will not be accessible either. We can estimate  $E_{* \text{virial}}$ directly from   
$E_{* \text{virial}} \approx E_{\text{virial}}/(1.5 N a^{1/3}) \approx  E_{\text{virial}}/|W|
\approx -0.33$ on the $T_{*+}$ branch of the two real solutions of the quartic
equation.  One can also obtain a less direct but similar estimate of $E_{*
\text{virial}}$ from the temperature $T_{* \text{virial}}$ of a cluster
modeled as a polytrope. To estimate the maximum kinetic energy, we note that if
the kinetic energy is twice the potential energy, the velocities in the unbound region will be about twice
the virial velocities that the region would have had if it were
virialized. These would be about the escape velocity, giving  $E_{*\text{max}} \approx 1$. More precise values of
these limits will follow from the solutions for $g(E_*)$ and the
binding probabilities. In Figure \ref{xt}, we show the regions of these
different solutions. Two of the four solutions of the quartic equation become
complex for $E_{*} \leq -0.39$ which is where the switch from the
$T_{*+}$ to the $T_{*-}$ real branches occurs. For $T_{*} >1$, the
total energy is positive and tends towards the perfect gas limit. With
these solutions for $T_* (E_*)$, we obtain the density of states
\begin{equation}
\begin{split} \label{e:ge1}
g[E_{*}(T_{*})] &= \dd{\Omega}{E_*} =  \dd{\Omega}{T_*} \dd{T_*}{E_*} \\
&=  \frac{3N}{2} e^{-N \ln N + 5N/2}~(1 +
T_{*}^{-3})^{N}~e^{-3N/(1+T_{*}^3)}~T_{*}^{(3N/2) -1} 
\end{split}
\end{equation}
represented here in terms of $T_*$ for simplicity, and setting
\begin{equation} \label{e:norm2}
\l(\frac{2 \pi m a^{1/3} V^{2/3}}{\Lambda^2}\r) = 1~.
\end{equation}
From $\eqref{e:ge1}$, we immediately see that there is a smooth
transition around $T_* = 1$, where the total energy becomes
negative. As $T_* \to 0$, the density of states becomes proportional to
$T_{*}^{-3N/2}$. This infinite number of negative energy states describes
a singularity. To reach this singularity, the motions must be
dominated by radial infall with negligible random velocities and
therefore negligible temperature. In this regime, the
quasi-equilibrium nature of the theory breaks down.

\begin{center}
\begin{figure}[!bth]
\plotone{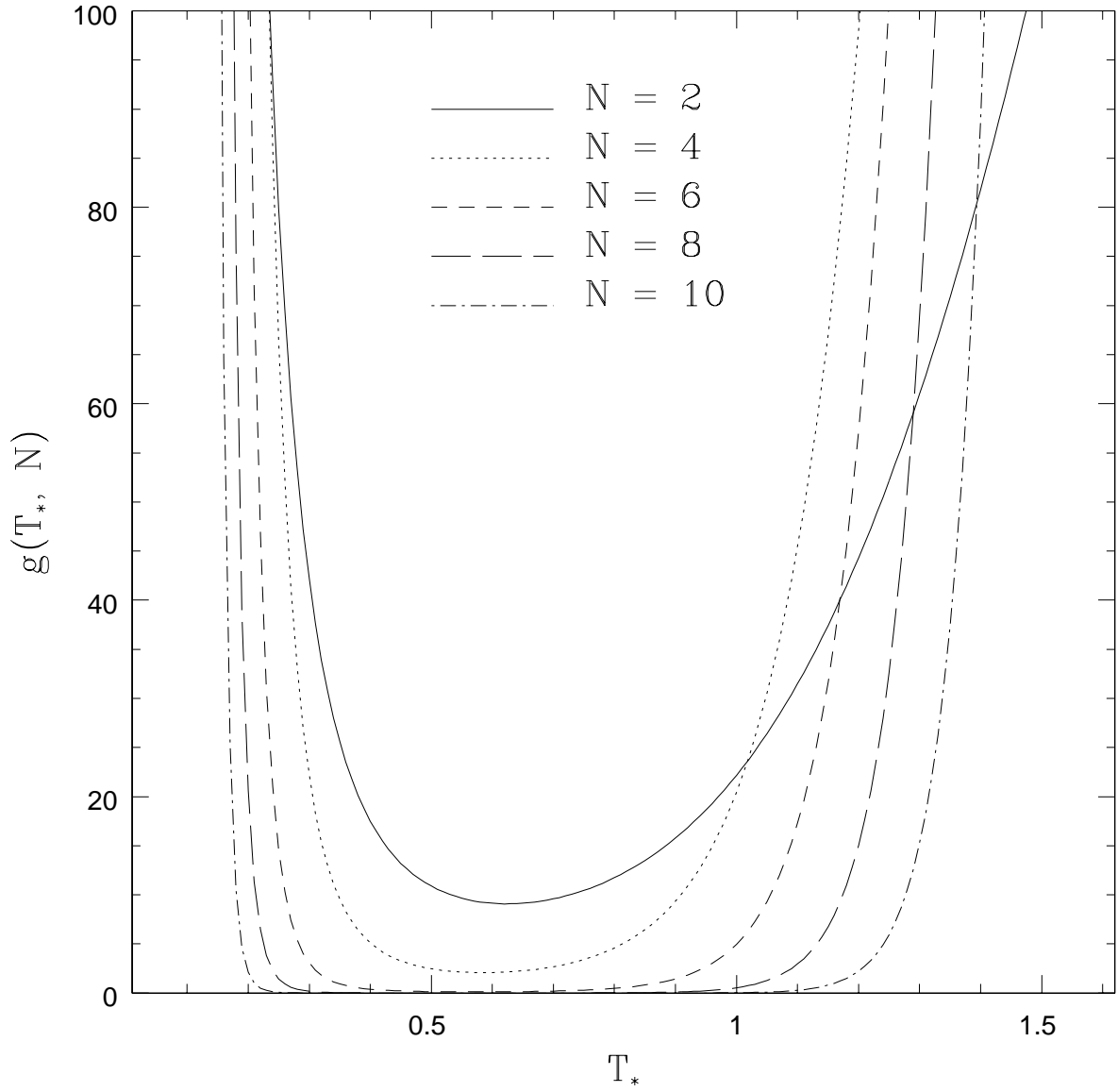}
\caption{The density of energy states, $g[E_{*} (T_{*})]$ for various values
of $N$.}  
\label{gfactor}
\end{figure}
\end{center}

\begin{center}
\begin{figure}[!bth]
\plotone{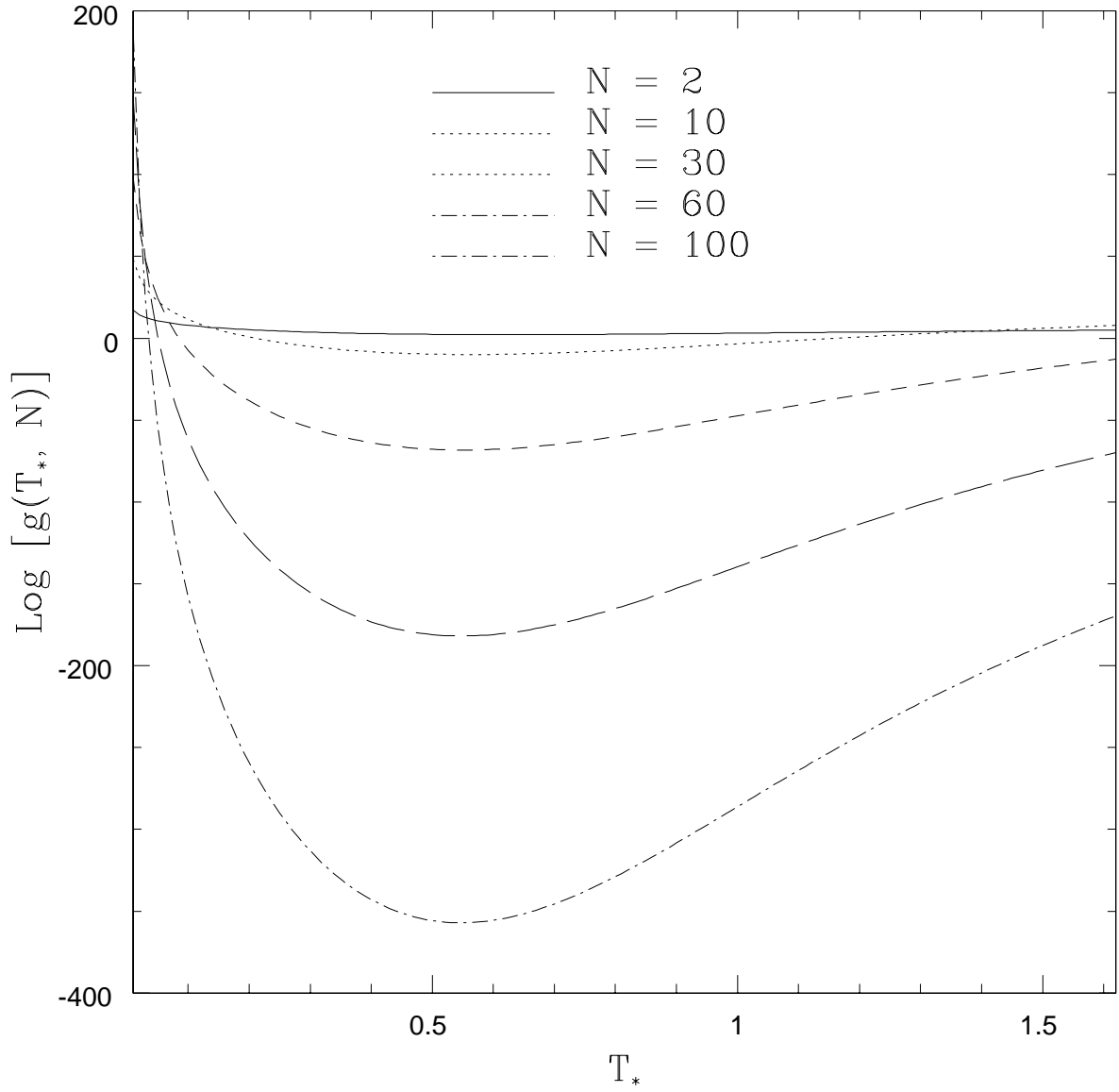}
\caption{The logarithm of the density of energy states for various values of
$N$. The curves all reach a minimum point at
$T_{*}$=0.54.}  
\label{logg}
\end{figure}
\end{center}

Figures \ref{gfactor} and \ref{logg} illustrate $g(T_*)$ as a function
of $T_{*}$, which is single valued unlike $E_{*}$, and also for a
range of $N$. It is clear 
from the linear plots in Figure 4, and from the logarithmic plots in
Figure 5 for a much greater range of $N$, that the $g[T_{*} (E_{*})]$
have infinite maxima both for the perfect gas and for gravitating
systems of galaxies as $T_{*} \to 0$. In the latter limit, the random
velocity dispersion $\langle v^2 \rangle \to 0$, or the gravitational energy
$\sim a^{1/3} \to \infty$, and the galaxies in a cell collapse into a
singularity having infinite entropy. The dynamical timescale for such
a collapse is so short that these systems cannot be in
quasi-equilibrium, and such states are inaccessible to the statistical
thermodynamic description. Moreover, they are also inaccessible
because they are not physically realistic. In any collapse toward a
singularity, it is very unlikely that all the radially infalling
orbits will remain in phase and reach the centre
simultaneously. Instead, they perturb each other, introducing more
random velocities which build up until the contracted system reaches
virial quasi-equilibrium which then evolves on a much longer
timescale. As shown in $\S 3.2$, this occurs around $T_{*} \approx
0.1$. Between this value and $T_{*} \approx 1.3$ the density of energy
states has a broad minimum corresponding to states which have neither a high
gravitational entropy nor a high perfect gas entropy. The total number
of configurations of these states is constrained by their relatively
narrow range of energy $\Delta E_{*}$ for a given range of temperature
$\Delta T_{*}$ around the minimum of the $E_{*} (T_{*} )$ relation in
Figure \ref{xt}. Therefore these are much less probable low entropy
states. 

Figures \ref{gfactor} and \ref{logg} also show the somewhat anomalous
behaviour of the case $N=2$, an effect of the approximation $3N/2
~\approx~(3N/2)-1$. However, for $N \gtrsim 4$, the curves become more
regular and this approximation becomes more reasonable; even for
$N=2$, it is qualitatively accurate. For larger $N$, the transition to
large $g$ becomes sharper and occurs at both lower and higher values
of $T_{*}$ than for small $N$. With $g(E_{*})$, we can now determine
$P(E_{*})$ for a canonical ensemble with each cell containing $N$
galaxies using equation $\eqref{e:prob1}$. The normalization
$\eqref{e:norm}$ cancels the contributions of $N$ in the chemical
potential and $Z_G$ although N still affects the form of $g[E_{*}
(T_{*})]$.

%%%%%%%%%%%%%%%%%%%%%%%%%%%%%%%%%%%%%%%%%%%%%%%%%%%%%%%%%%
\subsection{Binding and Virialization}
%%%%%%%%%%%%%%%%%%%%%%%%%%%%%%%%%%%%%%%%%%%%%%%%%%%%%%%%%%
The probability that $N$ galaxies in a cell are gravitationally bound
together is 
\begin{equation} \label{e:boundprob}
P_{\text{bound}} (N, T_{0*}) 
= \frac{ \int^{0}_{E_{* \text{min}}} P(E_*, N, T_{0*}) d E_* }{
\int^{E_{* \text{max}}}_{E_{*\text{min}}} P(E_*, N,
T_{0*}) d E_* }~.
\end{equation}
The number of bound states determines the number of unbound
states through the normalization $P_{\text{bound}} +
P_{\text{unbound}} = 1$. We have also computed $P_{\text{unbound}}$
separately as a check that the integrations satisfy this
normalization. 

As figures \ref{gfactor} and \ref{logg} show, the integrals depend somewhat
sensitively on the limits $E_{* \text{min}}$ and $E_{*
\text{max}}$ through their dependence on $T_{* \text{min}}$ and $T_{*
\text{max}}$ . We will explore this sensitivity around average values
of these limits. To estimate an average value for $E_{*\text{max}}$,
we note that for the $i$-th cell containing $N$ galaxies
\begin{equation} \label{e:virial1}
E_{* i} = \frac{K_i - |W_i|}{|\bar{W}|}
\end{equation}
where $|\bar{W}| = a^{1/3}$ is the average gravitational correlation
energy. If $|W_i| = |\bar{W}|$ in a typical state, then if the cell
were bound in virial equilibrium it would have $K_i = |\bar{W}|/2$. If
it were just bound, $K_i = |\bar{W}|$, and if it were sufficiently
unbound that the root mean square velocity were twice the virial root
mean square velocity, so that $K_i = 4 K_{\text{virial}} = 2 |
\bar{W}|$, then the configuration would expand to about twice its
size on a dynamical time scale. Such states are not in
quasi-equilibrium and would not be accessible to the statistical
mechanics or thermodynamics of the ensemble. Therefore we exclude
them. (In the statistical mechanics of perfect gases, this exclusion
is not necessary since particles of all energies are assumed to be in
equilibrium.) This suggests $E_{*\text{max}} \approx 1$ on the $T_{*+}$ branch. 

To estimate $E_{* \text{min}}$, it is more straightforward to work
with $T_{* \text{min}}$, which is a positive single-valued quantity,
and then find $E_{*\text{min}}$ through equation $\eqref{e:Estareqn}$.
Using equations $\eqref{e:a1}$ and $\eqref{e:Tstar}$ and the relation
between temperature and kinetic energy, gives 
\begin{equation} \label{e:qscond}
T_{*} = \frac{\bar{n}^{2/3} \langle v^2 \rangle}{3 (3/2)^{1/3}
\zeta^{1/3} G \bar{\rho}} = \frac{0.0881}{\zeta^{1/3}} \frac{\langle
v^2 \rangle \tau^2}{ \langle r_1 \rangle^2}~.
\end{equation}  
Here we have used the relation (cf equation 33.22 in Saslaw 1987)
connecting the average uniform number density $\bar{n}$ to the average
seperation $\langle r_1 \rangle$ between galaxies: $\bar{n}^{1/3} =
0.55/\langle r_1 \rangle$. The dynamical timescale is $\tau \equiv (G
\bar{\rho})^{-1/2}$, and $\zeta^{1/3}$ is usually unity or slightly
less from equation $\eqref{e:zeta}$. Although the minimum value of the
velocity dispersion for quasi-equilibrium to apply is not precisely
defined, it is necessary that $\langle v^2 \rangle \tau^2/\langle r_1
\rangle^2 \gtrsim 1$ so that the configuration does not collapse on a
dynamical time scale. This suggests $T_{* \text{min}} \approx 0.1$
corresponding to $E_{* \text{min}} \approx -0.1$ on the $T_{*-}$ branch.  

\begin{center}
\begin{figure}[!bth]
\plotone{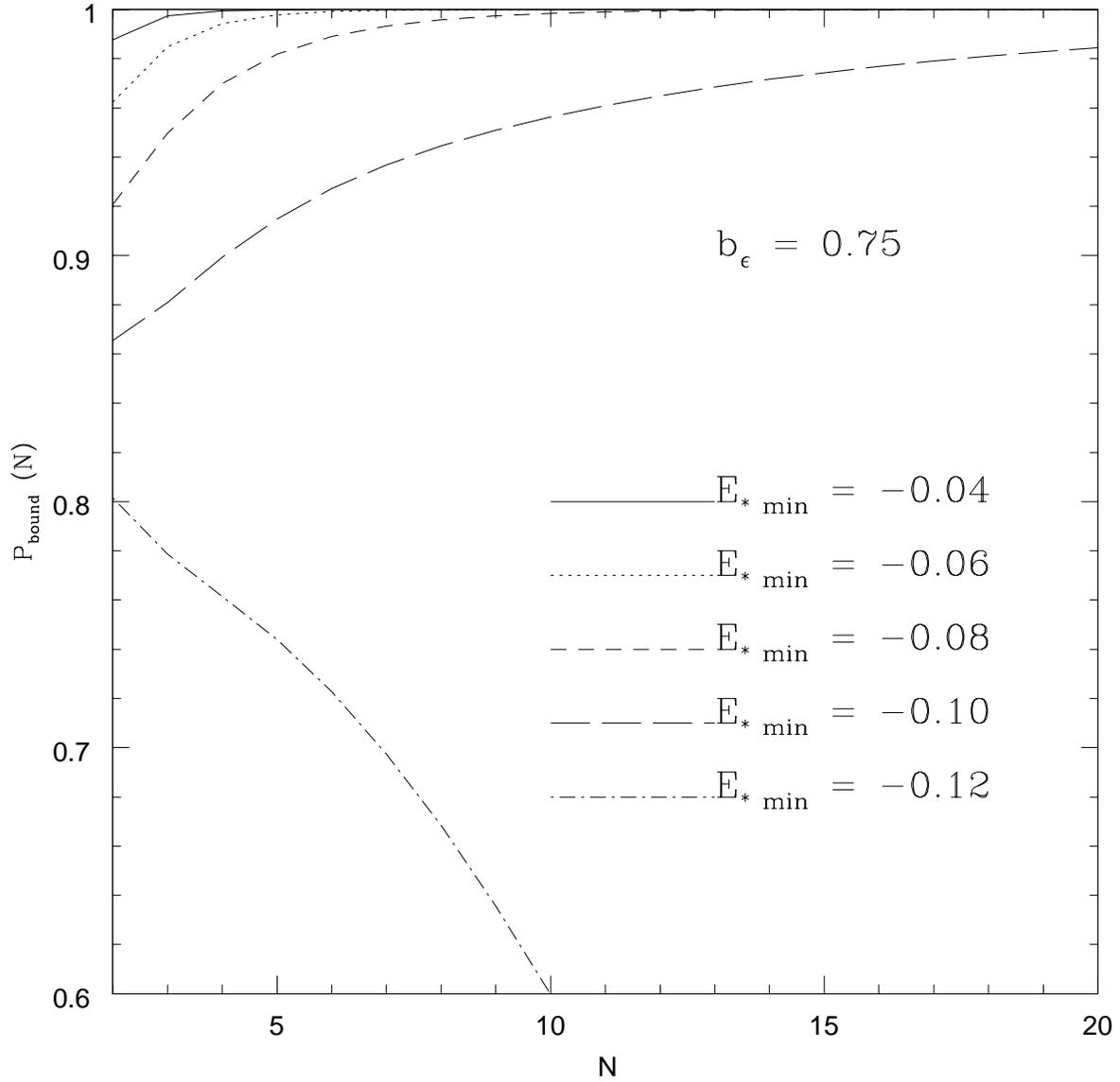}
\caption{The bound probabilities for $-0.12 \leq E_{* min} \leq
-0.04$. $E_{*min} = -0.1$ is the quasi-equilibrium limit. }  
\label{pbound}
\end{figure}
\end{center}

To check these estimates, we can compute $P_{\text{bound}} (N, E_{*
\text{min}} )$ from equation $\eqref{e:boundprob}$ for a range of
$E_{* \text{min}}$. The binding probability will also depend on
$b_{\epsilon}$ through its relation to $T_{0*}$ in equation
$\eqref{e:bepsilon}$. Figure \ref{pbound} shows the results for $-0.12
\leq E_{* \text{min}} \leq -0.04$ and $b_{\epsilon} = 0.75$. There is a
bifurcation at $E_{* \text{min}}=-0.10$ between bound probabilities which
increase with $N$ and those which decrease. We would expect that the
more galaxies there are in a cell of given volume, the more likely
they are to form a gravitationally bound group. Therefore functions
$P_{\text{bound}} (N)$ which decrease with increasing $N$ are
unphysical and, in Figure 6, only those cases with increasing
$P_{\text{bound}} (N)$ are valid. These unphysical cases have
$E_{* \text{min}} < -0.10$. However, $E_{*\text{min}} \approx -0.1$ is
also the smallest value from equation $\eqref{e:qscond}$ which is
consistent with quasi-equilibrium. Both arguments therefore suggest
that $E_{*\text{min}}\approx -0.10$ on the $T_{*-}$ branch of the $E_{*}
(T_{*})$ relation in Figure \ref{xt}, is the physical lower limit of the
probability integrals for $P_{\text{bound}} (N)$. 

\begin{center}
\begin{figure}[!bth]
\plotone{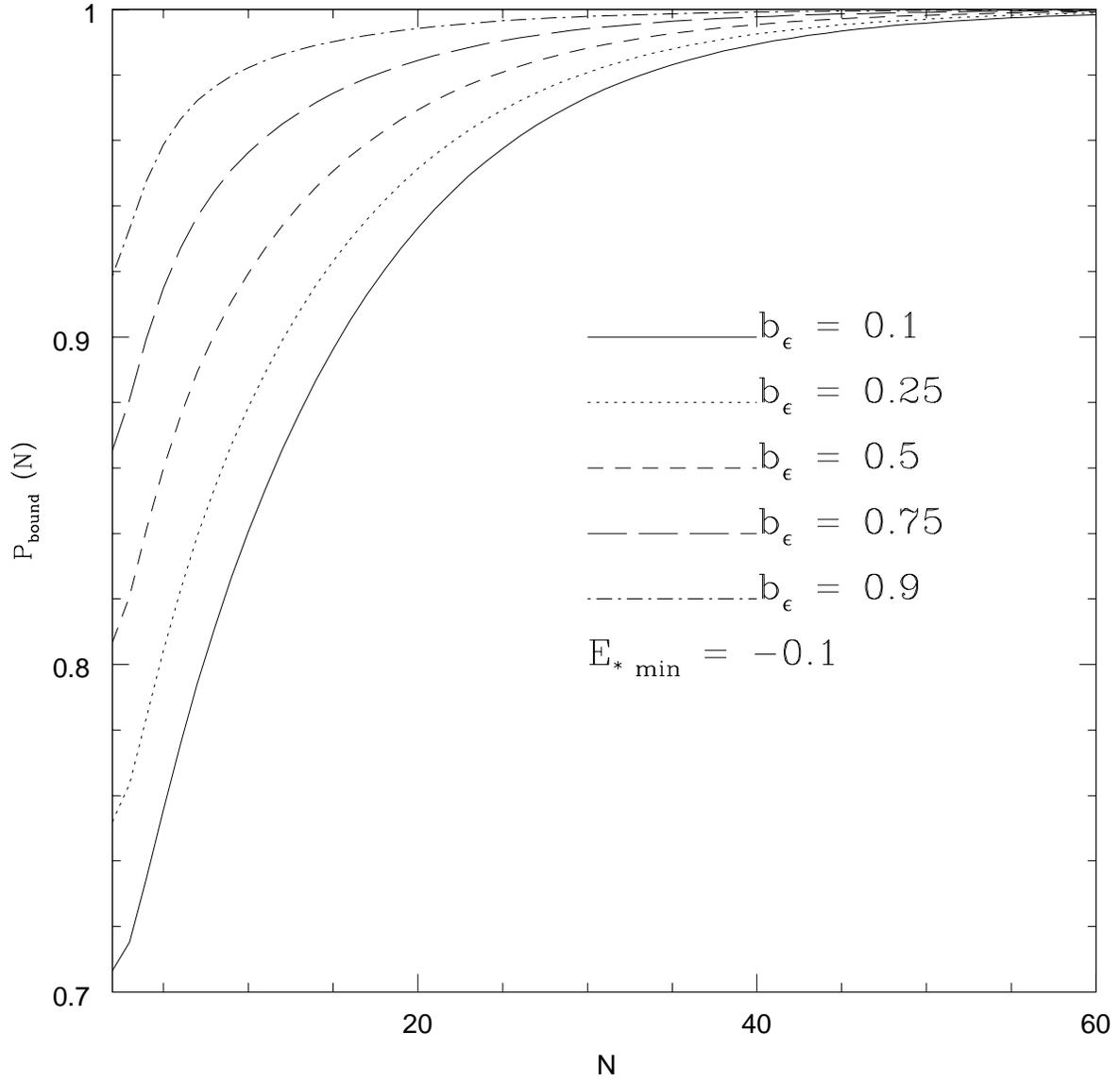}
\caption{The effect of $b_{\epsilon}$ on the bound probability. As
$b_{\epsilon}$ increases, the bound probability of clusters with
$N \geq 2$ increases, indicating greater efficiency in gravitational
binding.}  
\label{pbound2}
\end{figure}
\end{center}

With $E_{* \text{min}} = -0.1$ on the $T_{*-}$ branch, Figure \ref{pbound2}
shows the effect of $b_{\epsilon}$ on the binding probabilities for different $N$. 
Note that $\bar{N}$ enters through $a^{1/3}$ in equations
$\eqref{e:a1}$-$\eqref{e:Tstar}$ and the normalization in equation
$\eqref{e:norm2}$. As $b_{\epsilon}$ increases from $0.1$ for a nearly perfect gas to $0.9$
for a highly clustered gravitational system, the binding
probabilities increase significantly. These probabilities are
high even for relatively small values of $b_{\epsilon}$ which illustrates
the efficiency of gravitational clustering. Dynamically, positive
density perturbations expand more slowly than the surrounding 
universe even if they are linear, and eventually they become bound.  
In a statistical mechanical
description, these bound states have many more possible velocity and
density configurations than the corresponding unbound states. 
 
\begin{center}
\begin{figure}[!bth]
\plotone{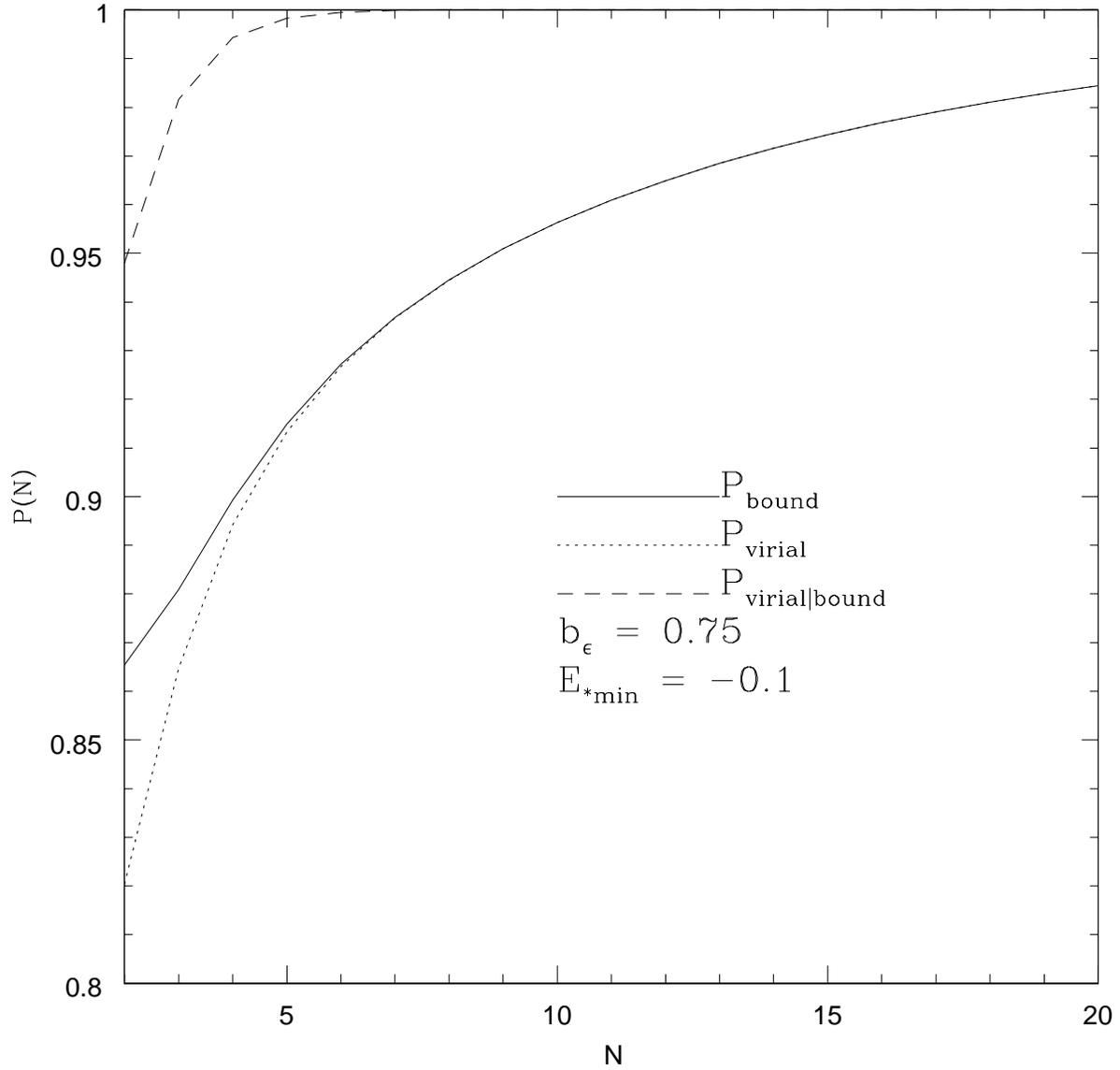}
\caption{The probabilities of bound and virialized 
clusters with $N \geq 2$ galaxies. The conditional probability that a bound
cluster is virialized is also plotted. Values of $b_{\epsilon}
= 0.75$ and $E_{*min} = -0.1$ are used.}  
\label{pbound3}
\end{figure}
\end{center}

One range of bound energy states is of special interest. These are the
approximately virialized states around $E_{*} \approx  -0.33$ on the $T_{*+}$ branch,
discussed after equation $\eqref{e:norm}$, which is $2 U_{*}/3$ in equation
$\eqref{e:Estar}$ with $U_{*} =-1/2$. Although the definition of the ``virial
range'' is somewhat arbitrary, here we take it to be $-0.39 \leq E_{*}
\leq -0.32$ on the $T_{*+}$ branch. Recall that for $E_{*} < -0.39$,
the $T_{*}(E_{*})$ relation becomes imaginary. Figure \ref{pbound3}
shows bound and virialized probabilities as well as the conditional probability
that a bound system of $N$ particles is in this virial range. Most bound states become virialized
and then evolve slowly so we should expect that many, including our
local group have been observed. 
%%%%%%%%%%%%%%%%%%%%%%%%%%%%%%%%%%%%%%%%%%%%%%%%%%%%%%%%%%%%
\subsection{Comparison with Observed Groups of Galaxies} 
\begin{center}
\begin{figure}[!bth]
\plotone{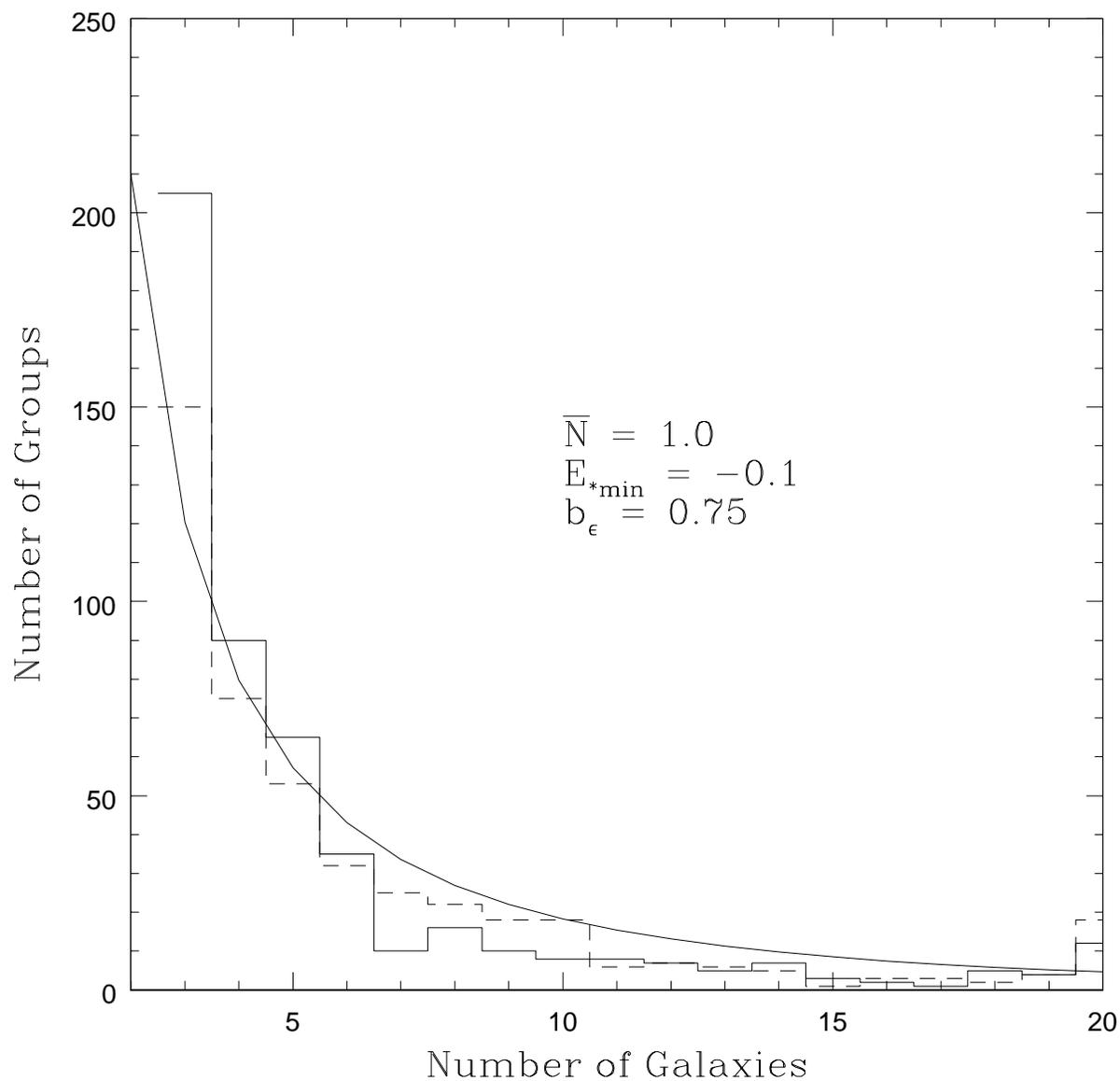}
\caption{Histograms are from Figure 2 in \cite{garcia} showing the
observed number distribution of nearby groups using the hierarchial
method (solid line) and the percolation method (dashed line). 
The solid curve with $b_{\epsilon} = 0.75$, $\overline{N}=1.0$ and
$E_{* min} = -0.1$ is our theoretical result.}  
\label{fb}
\end{figure}
\end{center}

%%%%%%%%%%%%%%%%%%%%%%%%%%%%%%%%%%%%%%%%%%%%%%%%%%%%%%%%%%
Identifying physically bound groups of observed galaxies is often
uncertain based on limited velocity and positional
information. Nevertheless, reasonable estimates of the numbers of such
groups have been made for the local universe where problems of
incomplete sampling have been minimized. These estimates also depend
on the criteria used to define a group. \cite{garcia} has explored
these issues in some detail and developed a catalog of 485 groups with
$N\geq 3$ members each in a selected sample of $6392$ nearby galaxies
with apparent magnitudes $B_0 \leq 14.0$ and radial velocities $\leq
5500$ km s$^{-1}$. Two procedures, one based on hierarchial
structures, the other based on percolation, are used to establish
group membership. From these, \cite{garcia} derived the number of
groups having $N \geq 3$ members  shown in Figure \ref{fb} as the solid
(hierarchial) and dashed (percolation) histograms taken from Garcia's
Figure 2. The moderate difference between these histograms is a
measure of the uncertainty of group membership. We use these results
for comparison with the expected probabilities of bound groups in
cosmological gravitational many-body clustering. 

To calculate the theoretical numbers of groups at the present time
(using the observed $b_{\epsilon}=0.75$ so that $T_{0*} = 3^{-1/3} =
0.70$), we multiply $f(N)$ from equation $\eqref{e:GQED1}$ by
$P_{\text{bound}}(N, T_{0*})$ from equation $\eqref{e:boundprob}$ and
normalize the total number to the 485 groups in Garcia's
catalog. Figure \ref{fb} shows the 
results with $E_{* \text{min}} = -0.10$ for the solid line. We have
also examined the results of varying $E_{*\text{min}}$ between -0.04
and -0.1. This range of $E_{* \text{min}}$ does not affect the
probabilities significantly. 

Overall there is reasonable agreement between the theoretical estimate
of the number of bound groups and the observed estimate for $3 \leq N
\leq 10$, which includes most of the sample. For $10 \leq N \leq 17$ the
predicted number is higher, while for $N=18, 19$, it agrees again with
the observations. The causes of the discrepancy are difficult to
determine, but may be related to the sensitive dependence of the numbers
in more populated groups on the magnitude cutoff. The theory assumes
that all galaxies have the same mass and luminosity for
simplicity and this could be improved. Larger catalogues, 
complete to fainter limiting apparent magnitudes will soon become
available and it will be interesting to see if they improve the
agreement.  

%%%%%%%%%%%%%%%%%%%%%%%%%%%%%%%%%%%%%%%%%%%%%%%%%%%%%%%%%%%%%%%%%
\section{THE COOL LOCAL HUBBLE FLOW}
%%%%%%%%%%%%%%%%%%%%%%%%%%%%%%%%%%%%%%%%%%%%%%%%%%%%%%%%%%%%%%%%%
We can apply these results to the long-standing problem of the
small peculiar velocity dispersion, $\lesssim 60$ km s$^{-1}$, of
galaxies in our local group \citep{vanden1, vanden2, ekholm}.
Consequently, the relatively small mass of the local group, ($\sim 2 
\pm 0.5 \times 10^{12} M_{\odot}$) does not significantly perturb the linear Hubble
flow beyond about 1.5 Mpc \citep{sandage1, sandage2, ekholm}. As
\cite{sandage3} has emphasized ``... the explanation of 
why the local expansion field {\it is} so noiseless remains a
mystery.''  The dynamics of the local flow and more general catalogs of
peculiar velocities have been explored  
by many authors \citep{groth, peebles, burstein, governato, willick1, willick2,
hoffman, nagamine}. 

Previously, there have been three main types of explanation for the cool
local Hubble flow. One, pioneered by \cite{kahn} and subsequently
often discussed, involves specific detailed models of the dynamics
of the formation of the local group. On the other hand,
there may be more generic reasons for low peculiar
velocities. \cite{sandage1} suggested two possibilities: a very low
density universe, or a high density universe dominated by a uniform
component of matter. In both these cases, gravitational galaxy
clustering would usually be relatively weak if it started from weakly
clustered initial conditions. The low density universe is not
generally considered likely at present, but the uniform high density
possibility has recently been revived by \cite{barychev} in terms of a
dark matter component suggested by a large positive value of the
cosmological constant.  

In explorations of these or other explanations, the fundamental question
is whether our weakly clustered local group is unusual enough to
demand a specific dynamical history, or whether it is quite
commonplace. Therefore we need to calculate the probability that a
small weakly bound group such as our
local group can form. Although our local group contains about three
dozen galaxies in a volume of about one megaparsec radius, most of its
mass and energy are dominated by Andromeda and the Milky way (including
their dark matter haloes). Most of the other galaxies are dwarfs or
satellites. This makes us a group with effectively $N=2$, or perhaps
$N=3$ major galaxies if all the smaller galaxies are equivalent to one
large galaxy.  

From Figure 6, at least 86$\%$ of all such groups are bound. Thus the
probability of a bound group such as our local cluster is
approximately $\gtrsim 0.86~f(N, b_{\epsilon})$. Using equation
$\eqref{e:GQED1}$ for $f(N, b_{\epsilon})$ with $\overline{N} = 1$ and $b_{\epsilon} = 0.75$ gives lower
limits to the probability that there are bound groups with $N$ effective massive galaxies 
within a radius of one megaparsec from an arbitrary point in space. These limits are $0.033$ 
and $0.019$ using effective values of $N=2$ or $N=3$ respectively for our local cluster. Thus 
within a radius of 10 Mpc, we would expect more than about $(4 \pi / 3)(1000)(0.033) = 135$ for
$N=2$ or 80 for $N=3$ groups such as our
own whose masses are too small to perturb their local Hubble flow
strongly. From equation $\eqref{e:GQED1}$, there would also be bound groups of greater mass which would 
perturb the Hubble flow more strongly, but such massive groups would be less numerous. 
Low mass bound groups like ours should be separated on average by $\sim 10(80)^{-1/3} \approx 2.3$ Mpc. 
So if they have a radius of $\sim 1$ Mpc and are randomly placed there is about
a 10$\%$ probability that a galaxy at a random position of space will belong to
such a group. Many galaxies will belong to even lower mass groups. This is also
a lower limit since most groups and galaxies are clustered rather than located
randomly. Therefore there is at least a 10$\%$ chance that an astronomer in an
arbitrary galaxy would belong to a group which does not seriously disturb the
surrounding cosmic expansion. This seems to us a high enough probability to
remove most of the mystery of the cool local Hubble flow.

%%%%%%%%%%%%%%%%%%%%%%%%%%%%%%%%%%%%%%%%%%%%%%%%%%%%%%%%%%%%%%%%%%%
\section{SUMMARY AND DISCUSSION}
%%%%%%%%%%%%%%%%%%%%%%%%%%%%%%%%%%%%%%%%%%%%%%%%%%%%%%%%%%%%%%%%%%%%
Distribution functions for peculiar velocities and gravitational
binding probabilities among galaxies provide new insights into their
clustering. These complement the more often analyzed distributions and
correlations of galaxy positions. Galaxy haloes, for example,
generally affect velocities more strongly than they affect
positions. Haloes soften the gravitational forces and inhibit the high
velocities associated with strong clustering. Strong clustering can
still develop at these lower velocities, it just takes longer.

Gravitational statistical mechanics is proving to be a powerful method for
calculating these distributions. It applies not only to the cosmological
many-body problem, but also to finite systems \citep{vega}, although since
finite systems have different symmetries they have a different thermodynamic
limit than our infinite system. In our cosmological case, the statistical
mechanical derivation of the peculiar velocity distribution function does not require assuming that average local
kinetic energy fluctuations are proportional to average local
potential energy fluctuations. We show that this assumption is a
direct consequence of the partition function. Moreover our new
derivation generalizes previous results by including possible haloes
around galaxies. The currently observed peculiar velocity distribution
function appears most consistent with each massive galaxy usually being
surrounded by a single individual halo rather than many galaxies sharing a large
common halo. Future extensive and accurate radial peculiar velocity
observations can quantify this result more precisely. 

The density of energy states for a system of gravitationally
interacting galaxies in the expanding universe is a fundamental
quantity. Usually the classical density of energy states can be
calculated in detail only for non-interacting systens, but our
cosmological gravitating system turns out to be a rare exception. 
Our calculation of this density of states makes it possible
to find the probability that a group of $N \geq 2$ galaxies is
gravitationally bound and virialized. This calculated probability
agrees well with the observed probability distribution for finding
groups of $N$ galaxies within about 100 Mpc. The preponderance of
relatively low mass groups also provides a simple explanation for the
observation that most groups, such as our own, do not disturb the
Hubble flow beyond the group appreciably. 

These results also show that the efficiency of gravitational galaxy
clustering is very high. More than 85$\%$ of double systems and more
than 95$\%$ of groups with $\gtrsim 10$ massive galaxies are bound. Of these at
least 95$\%$ are virialized. This is consistent with earlier studies
\citep{saslaw1979, saslaw3} which found that the observed form of the
galaxy two-point correlation function leads to highly efficient
clustering. 

The binding probability increases rapidly as $N$ increases. This is
more dramatic for small values of $b_{\epsilon}$ than for large
values. For large $b_{\epsilon}$, nearly all groups are bound for $N
\geq 2$. Since $b_{\epsilon}$ increases as the universe expands (a
manifestation of entropy increase), this describes the evolving
formation of bound groups and clusters. At the present time, nearly
all clusters with $N \gtrsim 60$ massive galaxies are expected to be
bound and virialized. Even for moderate values of $b_{\epsilon}
\approx 0.25$, which occur at redshifts $\sim 2.5 - 5$ depending on
the details of the expansion rate \citep{saslaw2001}, we would expect
most groups to be bound and virialized.

The high efficiency of binding and virialization on relatively small scales
provides further insight into the quasi-equilibrium nature of galaxy clustering.
On small scales, virialized clusters remain unstable only over dynamical
{\it relaxation} timescales $\sim$ $N/ \sqrt{G \rho}$, which are generally
longer than a Hubble time. On large scales, the formation of new clusters from
linear perturbations takes at least a Hubble time. Therefore changes in $b$ 
and the global equations of state take at least a Hubble time for their 
instability. This exceeds the local dynamical {\it crossing} timescale,
$1/\sqrt{G \rho}$, and is thus consistent with the approximation for
quasi-equilibirum \citep{saslaw2}.

Since the probability that a group forms depends on its total energy,
and its energy depends on its shape as well as on its size and number
of massive galaxies, our results make it possible to calculate the
probability that stable or unstable structures such as filaments or
``great walls'' can occur. We will analyze this elsewhere. 

%%%%%%%%%%%%%%%%%%%%%%%%%%%%%%%%%%%%%%%%%%%%%%%%%%%%%%%%%%%%%%%%%%%%%%
\acknowledgements
%%%%%%%%%%%%%%%%%%%%%%%%%%%%%%%%%%%%%%%%%%%%%%%%%%%%%%%%%%%%%%%%%%
The authors thank Morton Roberts and Yuri Barychev for helpful
discussions about the cool local Hubble flow and groups of nearby
galaxies. We also thank Phil Chan, Alvaro Dominguez, Jose Gaite
and Loh Yen Lee for several discussions, and the referee for suggestions
which helped clarify the manuscript. BL is supported by an
International Fellowship from the Agency of Science, Technology and
Research (A-STAR), Singapore.  

%%%%%%%%%%%%%%%%%%%%%%%%%%%%%%%%%%%%%%%%
%Bibliography
%%%%%%%%%%%%%%%%%%%%%%%%%%%%%%%%%%%%%%%%

%%%%%%%%%%%%%%%%%%%%%%%%%%%%%%%%%%%%%%%%%%%%%%%%%%%%%%%%%%%%%%%%
%Diagrams
%%%%%%%%%%%%%%%%%%%%%%%%%%%%%%%%%%%%%%%%%%%%%%%%%%%%%%%%%%%%%%%

\clearpage

\end{document}